\documentclass[journal]{IEEEtran}
\IEEEoverridecommandlockouts
\usepackage{cite}
\usepackage{hyperref}
\usepackage{amsmath,amssymb,amsfonts}
\usepackage{amsmath}
\usepackage{amsthm}

\usepackage{graphicx}
\usepackage{textcomp}
\usepackage{xcolor}
\usepackage{graphicx}
\usepackage{float}
\usepackage{subfigure}
\usepackage{amsmath}
\usepackage{amsfonts,amssymb}
\usepackage{mathrsfs}
\usepackage{mathtools}
\usepackage{algorithm}
\usepackage{algorithmicx}
\usepackage{algpseudocode}
\usepackage{bm}
\usepackage{multirow}
\usepackage{array}
\newtheorem{theorem}{Theorem}
\newtheorem{corollary}{Corollary}
\makeatletter
\newcommand{\biggg}{\bBigg@{3}}
\newcommand{\Biggg}{\bBigg@{3.5}}
\makeatother
\usepackage{stfloats}

\bibliography{IEEEabrv}
\def\BibTeX{{\rm B\kern-.05em{\sc i\kern-.025em b}\kern-.08em
    T\kern-.1667em\lower.7ex\hbox{E}\kern-.125emX}}
\newtheorem{lemma}{Lemma}
\newtheorem{remark}{Remark}
\begin{document}

\title{A Unified Analytical Framework for the Asymptotic Average Mutual Information Over Finite Input Mixture Gamma Distributed Channels\\
}

\author{Chongjun~Ouyang, Sheng~Wu, Chunxiao~Jiang,~\IEEEmembership{Senior Member,~IEEE,}\\
Yuanwei~Liu,~\IEEEmembership{Senior Member,~IEEE,}
Julian~Cheng,~\IEEEmembership{Senior Member,~IEEE,}
and Hongwen~Yang
\thanks{C. Ouyang, S. Wu, and H. Yang are with the School of Information and Communication Engineering, Beijing University of Posts and Telecommunications, Beijing, 100876, China (e-mail: \{DragonAim, thuraya, yanghong\}@bupt.edu.cn).}
\thanks{C. Jiang  is with the Tsinghua Space Center, Tsinghua University, Beijing, 100084, China, and also with the Beijing National Research Center for Information Science and Technology, Beijing, 100084, China (e-mail: jchx@tsinghua.edu.cn).}
\thanks{Y. Liu is with the School of Electronic Engineering and Computer Science, Queen Mary University of London, London, E1 4NS, U.K. (e-mail: yuanwei.liu@qmul.ac.uk).}
\thanks{J. Cheng is with the School of Engineering, The University of British Columbia, Kelowna, BC V1V 1V7, Canada (email: julian.cheng@ubc.ca).}
}
\maketitle

\begin{abstract}
This paper establishes a unified analytical framework to study the asymptotic average mutual information (AMI) of mixture gamma (MG) distributed fading channels driven by finite input signals in the high signal-to-noise ratio (SNR) regime. It is found that the AMI converges to some constant as the average SNR increases and its rate of convergence (ROC) is determined by the coding gain and diversity order. Moreover, the derived results are used to investigate the asymptotic optimal power allocation policy of a bank of parallel fading channels having finite inputs. It is suggested that in the high SNR region, the sub-channel with a lower coding gain or/and diversity order should be allocated with more power. Finally, numerical results are provided to approve the theoretical analyses.
\end{abstract}

\begin{IEEEkeywords}
Asymptotic analysis, finite inputs, mixture gamma distribution, power allocation.
\end{IEEEkeywords}

\section{Introduction}
Performance analysis (PA) of wireless transmissions over fading channels can offer valuable guidelines for the design and optimization of practical communication systems, which has been a focus of research since the 1960s, {e.g.}, see \cite{Giannakis2003,Tellambura2011,Liu2017,Kong2019,Zhang2018} and the references therein. The entire PA procedure usually contains two steps \cite{Giannakis2003}. Specifically, the first step aims to derive an exact or approximate expression of the analyzed performance metric for performance evaluation purposes, whereas the second step focuses on the metric's asymptotic behaviors in the high signal-to-noise ratio (SNR) regime in order to unveil how fading parameters influence the system performance.

As an often-used performance metric, the average channel capacity (ACC) has been widely studied \cite{Tellambura2011,Liu2017,Kong2019,Zhang2018}, which is defined as the maximal average mutual information (AMI) of fading channels. From an information-theoretic perspective, the ACC describes the highest error-free transmission rate and it can be achieved by Gaussian distributed input signals. Yet, practical transmit signals are taken from discrete finite constellation alphabets, e.g., quadrature amplitude modulation (QAM), instead of Gaussian ones, which makes the ACC not attainable \cite{Guo2005,Wu2018,Wu2011}. Thus, analyzing the AMI achieved by finite inputs is more meaningful than analyzing the ACC achieved by Gaussian inputs.

Sadly, it's challenging to perform theoretical analyses on the AMI over finite input fading channels, because the mutual information lacks closed-form expressions. Hence, most existing works on finite inputs focus on constellation or power allocation policy design \cite{Wu2018}. So far, very little attention has been paid to the performance analysis of the AMI over finite input fading channels \cite{Ouyang2020,Ramos2014,Rodrigues2014,Lozano2008}. Particularly, novel expressions were derived in \cite{Ouyang2020} to approximate the AMI, which, thus, finished the first step of the PA of the AMI. As for the second step, the authors in \cite{Ramos2014,Rodrigues2014,Lozano2008} studied the high-SNR asymptotics of the AMI over conventional smale-scale fading models, such as Rayleigh and Rician ones. Yet, for other more complicated or generalized fading models, e.g., the $\eta$-$\mu$ model, the corresponding AMI's high-SNR asymptotic behaviors are still unknown, which hence motivates this work.

In this paper, we derive novel expressions to portray the AMI's high-SNR asymptotic behaviors over wireless channels subject to the mixture gamma (MG) distributed fading and finite inputs. We also leverage the derived results to explore the asymptotic optimal power allocation policy in a bank of parallel independent MG distributed fading channels driven by finite inputs. Based on \cite{Tellambura2011}, the mixture gamma distribution (MGD) model is a versatile tool that can characterize many well-known fading distributions, such as the Rayleigh, Rician, Nakagami-$m$, $\eta$-$\mu$, $\kappa$-$\mu$, and $\mathcal K_G$ ones; thus, our derived results yield more unification and generality than those in \cite{Ramos2014,Rodrigues2014,Lozano2008}. Yet, the asymptotic AMI over the non-MG distributed fading channels will be discussed in our future works.

\section{System Model}
In a single-input single-output fading channel, the received signal can be written as
\begin{equation}\label{basic_model}
y=\sqrt{\bar\gamma}h s+z,
\end{equation}
where $z\sim{\mathcal{CN}}\left(0,1\right)$ denotes the additive white Gaussian noise (AWGN); $h\in{\mathbb C}$ denotes the fading coefficient satisfying ${\mathbb E}\left\{\left|h\right|^2\right\}=1$; $s$ denotes the normalized transmitted symbol and ${\mathbb E}\left\{\left|s\right|^2\right\}=1$; $\bar\gamma$ denotes the average received SNR. Assume that $s$ is taken from a finite constellation alphabet $\mathcal X$ consisting of $M$ points, $\left\{x_i\right\}_{i=1}^{M}$, with equal probabilities. Besides, we consider $a=\left|h\right|^2$ follows the MG distribution with its probability density function (PDF) and cumulative distribution function (CDF), respectively, given by \cite{Tellambura2011}
\begin{align}
f\left(a\right)&=\sum\nolimits_{l=1}^{L}\alpha_{l}a^{\beta_{l}-1}{\rm e}^{-\zeta_{l}a},\quad a\geq0,\label{EQUATION6}\\
F\left(a\right)&=\sum\nolimits_{l=1}^{L}\alpha_{l}\zeta_{l}^{-\beta_{l}}\Upsilon\left(\beta_{l},\zeta_{l}a\right),\label{EQUATION7}
\end{align}
where $L$, $\left\{\alpha_l\right\}$, $\left\{\beta_l\right\}$, and $\left\{\zeta_l\right\}$ denote the fading parameters satisfying $\int_{0}^{\infty}f\left(a\right){\rm d}a=\sum_{l=1}^{L}\alpha_{l}\Gamma\left({\beta_{l}}\right){\zeta_{l}^{-\beta_{l}}}=1$; $\Gamma\left(x\right)\triangleq\int_{0}^{\infty}t^{x-1}{\rm e}^{{-t}}{\rm{d}}t$ is the gamma function \cite[eq. (8.310.1)]{b15}; $\Upsilon\left(s,t\right)\triangleq\int_{0}^{t}x^{s-1}{\rm e}^{{-x}}{\rm{d}}x$ is the lower incomplete gamma function \cite[eq. (8.350.1)]{b15}. By \cite{Tellambura2011}, the MGD was proposed as an alternative model to characterize the often-used fading distributions, such as Nakagami-$m$, $\eta$-$\mu$, $\kappa$-$\mu$, and $\mathcal K_G$ ones, and the resultant parameters including $L$, $\left\{\alpha_l\right\}$, $\left\{\beta_l\right\}$, and $\left\{\zeta_l\right\}$ present different forms for different original fading types.


\section{Asymptotic Average Mutual Information}
\label{sec3}
Before analyzing the average mutual information of the fading channel, we first discuss the input-output mutual information of a scalar Gaussian channel: $Y=\sqrt{\gamma}S+Z$, where $Z\sim{\mathcal {CN}}\left(0,1\right)$ denotes the AWGN, $S$ is taken from the equiprobable constellation alphabet $\mathcal X$, and $\gamma$ is the SNR. By \cite{Ouyang2020}, the mutual information of this channel is given by
\begin{equation}
\begin{split}
I_{M}^{\mathcal X}\left(\gamma\right)=&\log_2{{M}}-\frac{1}{M\pi}\sum\nolimits_{j=1}^{M}\int_{\mathbb C}{{\rm e}^{-\left|u-\sqrt{\gamma}x_j\right|^2}}\\
&\times{\log_2{\left(\sum\nolimits_{k=1}^{M}{\rm e}^{\left|u-\sqrt{\gamma}x_j\right|^2-\left|u-\sqrt{\gamma}x_k\right|^2}\right)}}{\rm d}u.
\end{split}
\end{equation}
Thus, the average mutual information can be written as
\begin{align}\label{AMI_Def}
{\bar{{\mathcal I}}}_M=\int_{0}^{\infty}f\left(a\right)I_{M}^{\mathcal X}\left(\bar\gamma a\right){\rm d}a
=\int_{0}^{\infty}I_{M}^{\mathcal X}\left(t\right){\rm d}F\left(\frac{t}{\bar\gamma}\right).
\end{align}
Plugging \eqref{EQUATION6} into \eqref{AMI_Def} gives
\begin{align}
{\bar{{\mathcal I}}}_M&=\sum\nolimits_{l=1}^{L}\int_{0}^{\infty}\alpha_{l}a^{\beta_{l}-1}{\rm e}^{-\zeta_{l}a}I_{M}^{\mathcal X}\left(\bar\gamma a\right){\rm d}a\\
&\approx\sum\nolimits_{l=1}^{L}\sum\nolimits_{i=1}^{N}{\alpha_{l}\varpi_i}{\zeta_{l}^{-\beta_l}}\tau_i^{\beta_l-1}I_{M}^{\mathcal X}\left({\bar\gamma\tau_i}{\zeta_{l}^{-1}}\right),
\label{AMI_Appr}
\end{align}
where \eqref{AMI_Appr} follows the Gauss–Laguerre quadrature rule \cite[eq. (25.4.45)]{b14}; $\left\{\varpi_i\right\}$ and $\left\{\tau_i\right\}$ denote the weight and abscissas factors of the Gauss–Laguerre integration; $N$ is a complexity-vs-accuracy tradeoff parameter. We note that another approximate expression of ${\bar{{\mathcal I}}}_M$ was derived in \cite{Ouyang2020}. Unfortunately, the asymptotic behavior of ${\bar{{\mathcal I}}}_M$ when $\bar\gamma\rightarrow\infty$ was not characterized therein. In view of this issue, we now intend to analyze the asymptotic AMI by setting $\bar\gamma$ as infinity.

\subsection{A Unified Analytical Framework}
To facilitate the derivation, we rewrite \eqref{AMI_Def} as
\begin{align}
&{\bar{{\mathcal I}}}_M=\left.I_{M}^{\mathcal X}\left(t\right)F\left(\frac{t}{\bar\gamma}\right)\right|_{0}^{\infty}-\int_{0}^{\infty}F\left(\frac{t}{\bar\gamma}\right){\rm d}I_{M}^{\mathcal X}\left(t\right)\\
=&\log_2{M}-\sum\nolimits_{l=1}^{L}\frac{\alpha_{l}}{\zeta_{l}^{\beta_{l}}}\int_{0}^{\infty}\Upsilon\left(\beta_{l},\frac{\zeta_{l}t}{\bar\gamma}\right){\textrm{mmse}}_{M}^{\mathcal X}\left(t\right){\rm d}t,\label{AMI_Asym_Basic}
\end{align}
where ${\textrm{mmse}}_{M}^{\mathcal X}\left(\gamma\right)=\frac{{\rm d}I_{M}^{\mathcal X}\left(\gamma\right)}{{\rm d}\gamma}$ is the minimum mean-square error (MMSE) in estimating $S$ by observing $Y$ \cite{Guo2005}. Define ${\bar{{\mathcal I}}}_{M}^{\infty}\triangleq\lim_{\bar\gamma\rightarrow\infty}{\bar{{\mathcal I}}}_{M}$. When $\bar\gamma\rightarrow\infty$, we obtain $\frac{1}{\bar\gamma}\rightarrow0$, which together with \eqref{AMI_Asym_Basic} and the fact of $\lim_{x\rightarrow0}\Upsilon\left(s,x\right)=\frac{x^s}{s}+{\mathcal O}\left(x^{s+1}\right)$ \cite[eq. (8.354.1)]{b14}, yields
\begin{equation}\label{AMI_Asym_More}
\begin{split}
{\bar{{\mathcal I}}}_{M}^{\infty}&=\log_2{M}\\
&-\sum\nolimits_{l=1}^{L}\left(\frac{\alpha_l\hat{\mathcal{M}}\left({\beta_l}\right)}{\beta_l\bar\gamma^{\beta_l}}+
{\mathcal O}\left(\frac{\hat{\mathcal{M}}\left({\beta_l}+1\right)}{\bar\gamma^{\beta_l+1}}\right)
\right).
\end{split}
\end{equation}
Here, $\hat{\mathcal{M}}\left({x}\right)\triangleq{\mathcal M}\left[{\textrm{mmse}}_{M}^{\mathcal X}\left(t\right);{x}+1\right]$, where ${\mathcal M}\left[p\left(t\right);z\right]\triangleq\int_{0}^{\infty}t^{z-1}p\left(t\right){\rm d}t$ denotes the Mellin transform of $p\left(t\right)$ \cite{b18}. 
Since ${\textrm{mmse}}_{M}^{\mathcal X}\left(t\right)>0$ ($t>0$) \cite{Guo2005}, $\hat{\mathcal{M}}\left(x\right)>0$ holds when $x>0$. Before further derivations, we introduce the following two preliminary lemmas.
\vspace{-5pt}
\begin{lemma}
\label{theorem1}
  If the constellation alphabet $\mathcal X=\left\{x_i\right\}_{i=1}^{M}$ has no accumulation point, namely $d_{{\mathcal X},{\min}}\triangleq\inf_{i\neq j}\left|x_i-x_j\right|>0$, then it has ${\text{mmse}}_{M}^{\mathcal X}\left(\gamma\right)\leq{\mathcal O}\left(\gamma^{-\frac{1}{2}}{\rm e}^{-\frac{\gamma}{8}d_{{\mathcal X},{\min}}^2}\right)$ \cite{Wu2011}.
\end{lemma}
\vspace{-5pt}
\vspace{-5pt}
\begin{lemma}
\label{theorem0}
  If $p\left(t\right)$ is ${\mathcal O}\left(t^a\right)$ as $t\rightarrow0^{+}$ and ${\mathcal O}\left(t^b\right)$ as $t\rightarrow+\infty$, then $\left|{\mathcal M}\left[p\left(t\right);z\right]\right|<\infty$ when $-a<z<-b$ \cite{b18}.
\end{lemma}
\vspace{-5pt}
We comment that this work focuses more on the constellation alphabet used in practical systems, for example $M$-QAM, and the condition of $d_{\mathcal X,\min}=\inf_{i\neq j}\left|x_i-x_j\right|>0$ can be generally satisfied. Hence, Lemma \ref{theorem1} can be directly used. Based on these preliminaries, we now establish the following theorem that provides a unified framework for analyzing the asymptotic AMI over MG distributed fading channels.
\vspace{-5pt}
\begin{theorem}
\label{theorem2}
  In the high SNR region, the asymptotic AMI over mixture gamma distributed fading channels satisfies
  \begin{align}\label{Asym_AMI_Case5}
    {\bar{{\mathcal I}}}_{M}^{\infty}=\log_2{M}-{G_{a,M}}{\bar\gamma^{-G_{d,M}}}+{\mathcal O}\left(\bar\gamma^{-G_{d,M}-1}\right),
  \end{align}
where ${G_{d,M}}=\beta_1>0$, $G_{a,M}=G_{a,M,r}\hat{\mathcal M}\left({G_{d,M}}\right)<\infty$, and $G_{a,M,r}=\beta_1^{-1}{\sum_{\beta_l=\beta_{1}}\alpha_l}$.
\end{theorem}
\vspace{-5pt}
\begin{IEEEproof}
Please see Appendix \ref{Append0}.
\end{IEEEproof}
As shown, $G_{a,M,r}$ and $G_{d,M}$ are determined by the fading parameters, whereas $G_{a,M}$ is determined by both the fading parameters and the structure of the adopted constellation alphabet. Based on \cite{Tellambura2011}, $\alpha_l>0$, which yields $G_{a,M,r}>0$ and thus $G_{a,M}>0$. Hence, the AMI converges to $\log_2{M}$ as $\bar\gamma\rightarrow\infty$ and its rate of convergence (ROC) is shaped by $G_{a,M}$ and $G_{d,M}$. Clearly, the rate of ${\bar{{\mathcal I}}}_{M}^{\infty}$ converging to $\log_2{M}$ equals the rate of ${G_{a,M}}{\bar\gamma^{-G_{d,M}}}$ converging to 0. Interestingly, ${G_{a,M}}{\bar\gamma^{-G_{d,M}}}$ follows the classical ``diversity order-coding gain'' pattern \cite{Giannakis2003} with its diversity order and coding gain given by $G_{d,M}$ and $G_{a,M}^{-1/{G_{d,M}}}$, respectively. Moreover, we find that the structure of the adopted constellation alphabet affects the ROC of ${\bar{{\mathcal I}}}_{M}^{\infty}$ via the Mellin transform of the MMSE function.

\subsection{Case Study}\label{SECTION3B}
To show the generality of the established analytical framework, we now intend to use Theorem \ref{theorem2} to discuss the asymptotic AMI of four special MG distributed fading channels.
\subsubsection{Nakagami-$m$ Fading}
For Nakagami-$m$ fading channels, $L=1$, $\alpha_1=\frac{m^m}{\Gamma\left(m\right)}$, $\beta_1=m$, and $\zeta_1={m}$ \cite{Tellambura2011}. Based on this and Theorem \ref{theorem2}, we arrive at the following corollary.
\vspace{-5pt}
\begin{corollary}
\label{corollary1}
For the asymptotic AMI over Nakagami-$m$ fading channels, its diversity order and coding gain are given by $m$ and $\left(\frac{m^{m-1}}{\Gamma\left(m\right)}\hat{\mathcal M}\left(m\right)\right)^{-1/m}$, respectively.
\end{corollary}
\vspace{-5pt}

\subsubsection{$\eta$-$\mu$ Fading}
According to \cite[eq. (13)]{Tellambura2011}, for $\eta$-$\mu$ fading channels, we have
\begin{equation}
\begin{split}
&L=\infty,~~\alpha_l=\psi\left(\theta_l,\beta_l,\zeta_l\right),~~\beta_l=2\left(\mu-1+l\right),\\
&\zeta_l={2\mu h},~~\theta_l=\frac{2\sqrt{\pi}\mu^{2\mu+2l-2}h^{\mu}H^{2l-2}}{\Gamma\left(\mu\right)\Gamma\left(l\right)\Gamma\left(\mu+l-\frac{1}{2}\right)},
\end{split}
\end{equation}
where $\psi\left(\theta_l,\beta_l,\zeta_l\right)\triangleq\frac{\theta_l}{\sum_{i=1}^{L}\theta_i\Gamma\left(\beta_i\right)\zeta_i^{-\beta_i}}$. Additionally, $\eta$-$\mu$ fading consists of two formats: $\left.\textrm{\lowercase\expandafter{\romannumeral1}}\right)$ Format \uppercase\expandafter{\romannumeral1}, $0<\eta<\infty$, $H=\left(\eta^{-1}-\eta\right)/4$, and $h=\left(2+\eta^{-1}+\eta\right)/4$; $\left.\textrm{\lowercase\expandafter{\romannumeral2}}\right)$ Format \uppercase\expandafter{\romannumeral2}, $-1<\eta<1$, $H = \eta/\left(1-\eta^2\right)$, and $h = 1/\left(1-\eta^2\right)$. We notice that $\beta_1<\beta_2<\cdots<\beta_L$ holds; thus, the following corollary can be found.
\vspace{-5pt}
\begin{corollary}
\label{corollary2}
For the asymptotic AMI over $\eta$-$\mu$ fading channels, its diversity order and coding gain can be written as $2\mu$ and $\left(\frac{\alpha_1}{2\mu}\hat{\mathcal M}\left(2\mu\right)\right)^{-1/2\mu}$, respectively.
\end{corollary}
\vspace{-5pt}
Additionally, ${\sum_{i=1}^{+\infty}\theta_i\Gamma\left(\beta_i\right)\zeta_i^{-\beta_i}}=\frac{h^{\mu}2^{2\mu}\mu^{2\mu}}{\left[\left({2\mu h}\right)^{2}-\left({2\mu H}\right)^2\right]^{\mu}}$ \cite[eq. (14)]{Ouyang2020}, which leads to $\alpha_1=\frac{2\sqrt{\pi}\mu^{2\mu}\left(h^2-H^2\right)^\mu}{\Gamma\left(\mu\right)\Gamma\left(\mu+\frac{1}{2}\right)}$.

\subsubsection{$\kappa$-$\mu$ Fading}
On the basis of \cite[eq. (19)]{Tellambura2011}, for $\kappa$-$\mu$ fading channels, we have
\begin{equation}
\begin{split}
&L=\infty,~~\alpha_l=\psi\left(\theta_l,\beta_l,\zeta_l\right),~~\beta_l=\mu-1+l,\\
&\zeta_l={\mu\left(1+\kappa\right)},~~\theta_l=\frac{\mu^{\mu+2l-2}\kappa^{l-1}\left(1+\kappa\right)^{\mu+l-1}}{{\rm e}^{\mu\kappa}\Gamma\left(\mu+l-1\right)\Gamma\left(l\right)},
\end{split}
\end{equation}
where $\kappa>0$ and $\mu>0$. By following the similar approach in the former part, we can obtain the following corollary.
\vspace{-5pt}
\begin{corollary}
\label{corollary3}
For the asymptotic AMI over $\kappa$-$\mu$ fading channels, its diversity order and coding gain can be written as $\mu$ and $\left(\frac{\alpha_1}{\mu}\hat{\mathcal M}\left(\mu\right)\right)^{-1/\mu}$, respectively.
\end{corollary}
\vspace{-5pt}
Based on \cite[eq. (14)]{Ouyang2020}, ${\sum_{i=1}^{+\infty}\theta_i\Gamma\left(\beta_i\right)\zeta_i^{-\beta_i}}=1$, which yields $\alpha_1=\frac{\mu^{\mu}\left(1+\kappa\right)^{\mu}}{{\rm e}^{\mu\kappa}\Gamma\left(\mu\right)}$. Last, but not least, we comment that Rician fading is a special case of the $\kappa$-$\mu$ fading when $\mu=1$ and $\kappa$ denotes the Rician factor \cite{Tellambura2011}.

\subsubsection{${\mathcal K}_G$ Fading}
According to \cite[eq. (9)]{Tellambura2011}, for ${\mathcal K}_G$ fading channels, we have
\begin{equation}
\begin{split}
&L=N,~~\alpha_l=\psi\left(\theta_l,\beta_l,\zeta_l\right),~~\beta_l=m,\\
&\zeta_l=\frac{km}{\tau_l},~~\theta_l=\frac{k^mm^m\varpi_l\tau_l^{k-m-1}}{\Gamma\left(m\right)\Gamma\left(k\right)},
\end{split}
\end{equation}
where $N>0$ is a complexity-vs-accuracy tradeoff parameter; $k>0$ and $m>0$ are fading coefficients of the ${\mathcal{K}}_{G}$ fading model; $\left\{\varpi_i\right\}$ and $\left\{\tau_i\right\}$ denote the weight and abscissas factors of the Gauss–Laguerre integration satisfying $\int_{0}^{\infty}t{\rm e}^{-t}{\rm d}t\approx\sum_{i=1}^{N}\varpi_i\tau_i$. The following corollary can be found.
\vspace{-5pt}
\begin{corollary}
\label{corollary4}
For the asymptotic AMI over ${\mathcal K}_G$ fading channels, its diversity order and coding gain can be written as $m$ and $\left(\frac{\sum_{l=1}^{N}\alpha_1}{m}\hat{\mathcal M}\left(m\right)\right)^{-1/m}$, respectively.
\end{corollary}
\vspace{-5pt}
Particularly, $\sum_ {l=1}^{N}\alpha_l=\frac{\sum_{l=1}^{N}\theta_l}{\sum_{i=1}^{N}\theta_i\Gamma\left(\beta_i\right)\zeta_i^{-\beta_i}}\overset{(a)}{\approx}\sum_{l=1}^{N}\theta_l$, where the step ``$(a)$'' is based on the statement in \cite[Section \uppercase\expandafter{\romannumeral4}]{Ouyang2020}, namely $\sum_{i=1}^{N}\theta_i\Gamma\left(\beta_i\right)\zeta_i^{-\beta_i}\approx1$.

Apart from the discussed four special cases, the established analytical framework also apply to other well-known fading distributions that serve as special or limiting cases of the MGD model, such as the fluctuating two-ray (FTR) \cite{Zhang2018}, $\alpha$-$\lambda$-$\eta$-$\mu$/gamma \cite{Hmood2020}, and Nakagami-lognormal (NL) \cite{Tellambura2011} ones. Last, but not least, for other more complex fading distributions whose PDFs cannot be explicitly written in terms of the MG distribution, one may obtain a MGD-based approximation of their PDFs aided with the expectation maximization (EM) algorithm \cite{Alhussein2014}. Yet, this is beyond the scope of this paper.

\section{Application: Asymptotic Optimal Power Allocation for Parallel Fading Channels}
\label{SECTION4}
In the sequel, we provide an example to show the application of the derived results, where \eqref{Asym_AMI_Case5} is exploited to discuss the optimal power allocation policy of a bank of parallel channels \cite{Lozano2006} in the high SNR regime. Particularly, in a system consisting of $K$ parallel independent fading sub-channels, the received signal in the $k$-th sub-channel can be expressed as
\begin{equation}
y_k=\sqrt{\bar\gamma}\sqrt{p_k}h_ks_k+z_k,~k\in{\mathcal K}\triangleq\left\{1,2,\cdots,K\right\},
\end{equation}
where $z_k\in{\mathcal{CN}}\left(0,1\right)$ denotes the AWGN; $s_k$ denotes the $k$-th sub-channel input taken from the equiprobable alphabet $\mathcal X=\left\{x_i\right\}_{i=1}^{M}$ and ${\mathbb E}\left\{\left|s_k\right|^2\right\}=1$; $h_k$ denotes the fading coefficient of the $k$-th sub-channel; the fading power $a_k=\left|h_k\right|^2$ follows the mixture gamma distribution with its PDF given by $f_k\left(x\right)$ and its mean given by ${\mathbb E}\left\{a_k\right\}=1$; ${\textbf p}\triangleq\left[p_1,p_2,\cdots,p_K\right]$ represents the vector of power allocation policy satisfying $\sum_{i=1}^{K}p_k=1$ and $p_k\geq0$; $\bar\gamma$ denotes the total available transmit power. We assume that $x_1,x_2,\cdots,x_K$, $z_1,z_2,\cdots,z_K$, and $h_1,h_2,\cdots,h_K$ are independent random variables. Besides, we assume that the receiver knows the instantaneous channel state information perfectly but the transmitter only knows the distribution of the sub-channel gains \cite{Lozano2008,Ramos2014}. Under this model, the power allocation problem can be formulated as \cite{Lozano2008,Ramos2014}
\begin{subequations}
\begin{align}
P_0:~\min_{\textbf p}&~{\mathcal{S}}_M^{\mathcal X}\left(\textbf p\right)=-\sum\nolimits_{k=1}^{K}{\mathbb E}\left\{I_M^{\mathcal X}\left(a_kp_k\bar\gamma \right)\right\}\\
{\textrm{s.t.}}&~\sum\nolimits_{k=1}^{K}p_k=1;~p_k\geq0,~\forall k\in{\mathcal K}.
\end{align}
\end{subequations}
Based on \cite{Guo2005}, $I_M^{\mathcal X}\left(x\right)$ is a concave function of $x$ and thus ${\mathcal{S}}_M^{\mathcal X}\left(\textbf p\right)$ is a convex function of $\textbf p$. Consequently, problem $P_0$ is a standard convex optimization problem. Furthermore, on the basis of \eqref{Asym_AMI_Case5}, when $\bar\gamma\rightarrow\infty$, we can write the asymptotic AMI of the $k$-th sub-channel as
\begin{align}\label{AMI_Asym_Final2_Appl}
{\mathbb E}\left\{{I}_M^{\mathcal X}\left(a_k\bar\gamma \right)\right\}=\log_2{M}-\frac{{A_{M,k}}}{\bar\gamma^{{D_{M,k}}}}+{\mathcal O}\left(\frac{1}{\bar\gamma^{{D_{M,k}}+1}}\right).
\end{align}
For clarity, let us denote the optimal solution of $P_0$ as ${\textbf p}^{\star}=\left[p_1^{\star},p_2^{\star},\cdots,p_K^{\star}\right]$ and $D_{\min}= \min_{k\in\mathcal K}D_{M,k}$. Leveraging the asymptotic expression in \eqref{AMI_Asym_Final2_Appl}, the following theorem can be found to capture the main results of the optimal power allocation policy in the high SNR region.
\vspace{-5pt}
\begin{theorem}
\label{lemma1}
  In the high SNR region, for $k\in\mathcal K$ that satisfies $D_{M,k}=D_{\min}$, we have
  \begin{align}
  p_k^{\star}={\left(\frac{A_{M,k}D_{\min}}{\nu{\bar\gamma^{D_{\min}}}}\right)}^{\frac{1}{D_{\min}+1}}+{\mathcal O}\left({\bar\gamma^{-1}}\right),~\bar\gamma\rightarrow\infty,\label{Asym_Power_Result2}
  \end{align}
  whereas for $k\in\mathcal K$ that satisfies $D_{M,k}>D_{\min}$, we have
  \begin{align}
    p_k^{\star}=&{\left(\frac{A_{M,k}D_{M,k}}{\nu{\bar\gamma^{D_{\min}}}}\right)}^{\frac{1}{D_{M,k}+1}}{\bar\gamma^{\frac{D_{\min}-D_{M,k}}{D_{M,k}+1}}}\nonumber\\
    &+{\mathcal O}\left({\bar\gamma^{\left(D_{\min}-D_{M,k}\right){\frac{D_{M,k}+2}{D_{M,k}+1}}-1}}\right),~\bar\gamma\rightarrow\infty.\label{Asym_Power_Result3}
  \end{align}
  In \eqref{Asym_Power_Result2} and \eqref{Asym_Power_Result3}, $\nu$ ensures that $\sum_{k=1}^{K}p_k^{\star}=1$, and $\nu$ satisfies
  \begin{align}
  \nu{\bar\gamma^{D_{\min}}}={\mathcal O}\left(1\right),\quad\bar\gamma\rightarrow\infty.
  \end{align}
\end{theorem}
\vspace{-5pt}
\begin{IEEEproof}
Please see Appendix \ref{Append1}.
\end{IEEEproof}
Notably, the power allocation policy put forth in Theorem \ref{lemma1} is still effective even though different constellation alphabets are utilized among these sub-channels. Besides, Theorem \ref{lemma1} encompasses the results in \cite{Lozano2008,Ramos2014} as its special cases when the MGD model is degenerated to Rayleigh and Rician models, respectively. Interestingly, based on Theorem \ref{lemma1}, as $\bar\gamma\rightarrow\infty$, $p_k^{\star}$ gradually collapses to but never equals zero for $D_{M,k}>D_{\min}$ whereas $p_k^{\star}\rightarrow{\left(\frac{A_{M,k}D_{\min}}{\nu{\bar\gamma^{D_{\min}}}}\right)}^{\frac{1}{D_{\min}+1}}$ for $D_{M,k}=D_{\min}$. According to this and \eqref{Asym_Power_Result2}, we can obtain
\begin{align}
\sum\nolimits_{D_{M,k}=D_{\min}}{\left(\frac{A_{M,k}D_{\min}}{\nu{\bar\gamma^{D_{\min}}}}\right)}^{\frac{1}{D_{\min}+1}}=1,~\bar\gamma\rightarrow\infty,
\end{align}
and it follows, for $D_{M,k}=D_{\min}$, that
\begin{align}
p_k^{\star}=
\frac{{A_{M,k}}^{\frac{1}{D_{\min}+1}}}{\sum_{D_{M,l}=D_{\min}}{A_{M,l}}^{\frac{1}{D_{\min}+1}}},~\bar\gamma\rightarrow\infty.
\end{align}
Based on this, we proceed with the following remark.
\vspace{-5pt}
\begin{remark}\label{Remark1}
In the high SNR regime, most power will be allocated to the sub-channels with the lowest diversity order.
\end{remark}
\vspace{-5pt}
The results in Remark \ref{Remark1} can be explained from two aspects. One is that the AMI of each sub-channel tends to its maximum, $\log_2{M}$, as $\bar\gamma$ increases, the other is that in contrast to the sub-channel with a higher diversity order, the sub-channel with a lower diversity order requires more power to approach its maximal AMI. Thus, to maximize the AMI of each sub-channel in the high SNR region, more power should be allocated to the sub-channels with lower diversity orders. Finally, consider a special case when all the sub-channels possess the same diversity order, namely $D_{M,1}=D_{M,2}=\cdots=D_{M,K}=D_{\text{c}}$. By following similar derivations as described in Appendix \ref{Append1}, the optimal power allocation policy in the high SNR region can be expressed as
\begin{align}\label{Power_Allocate_Prop2}
p_k^{\star}=\frac{{A_{M,k}}^{\frac{1}{D_{\text{c}}+1}}}{\sum_{l=1}^{K}{A_{M,l}}^{\frac{1}{D_{\text{c}}+1}}}+{\mathcal O}\left(\bar\gamma^{-1}\right),~\forall k,~\bar\gamma\rightarrow\infty,
\end{align}
which suggests that the smaller the coding gain, namely $A_{M,k}^{-1/D_{\text c}}$, the higher power the sub-channel can be allocated. The reason for this is similar as that for Remark \ref{Remark1}.

\section{Numerical Results}
\label{sec4}
\begin{figure}[!t]
\vspace{-10pt}
    \centering
    \subfigbottomskip=-5pt
	\subfigcapskip=-5pt
\setlength{\abovecaptionskip}{5pt}
    \subfigure[Nakagami-$m$, $m=2$.]
    {
        \includegraphics[height=0.185\textwidth]{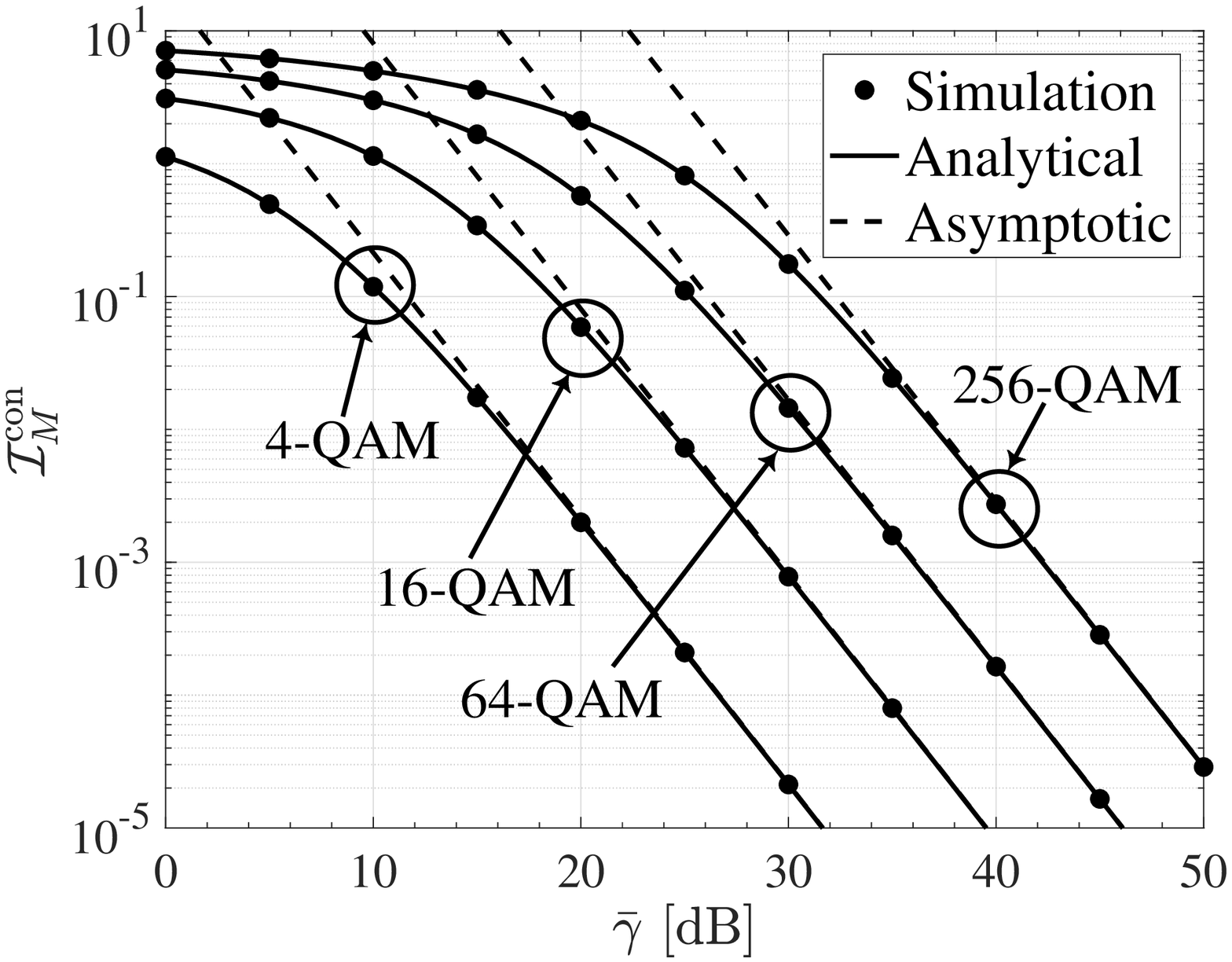}
	   \label{fig1a}	
    }
    \hspace{-23pt}
   \subfigure[$\eta$-$\mu$ Format \textrm{\uppercase\expandafter{\romannumeral1}}, $\eta=4$, $\mu=1$.]
    {
        \includegraphics[height=0.185\textwidth]{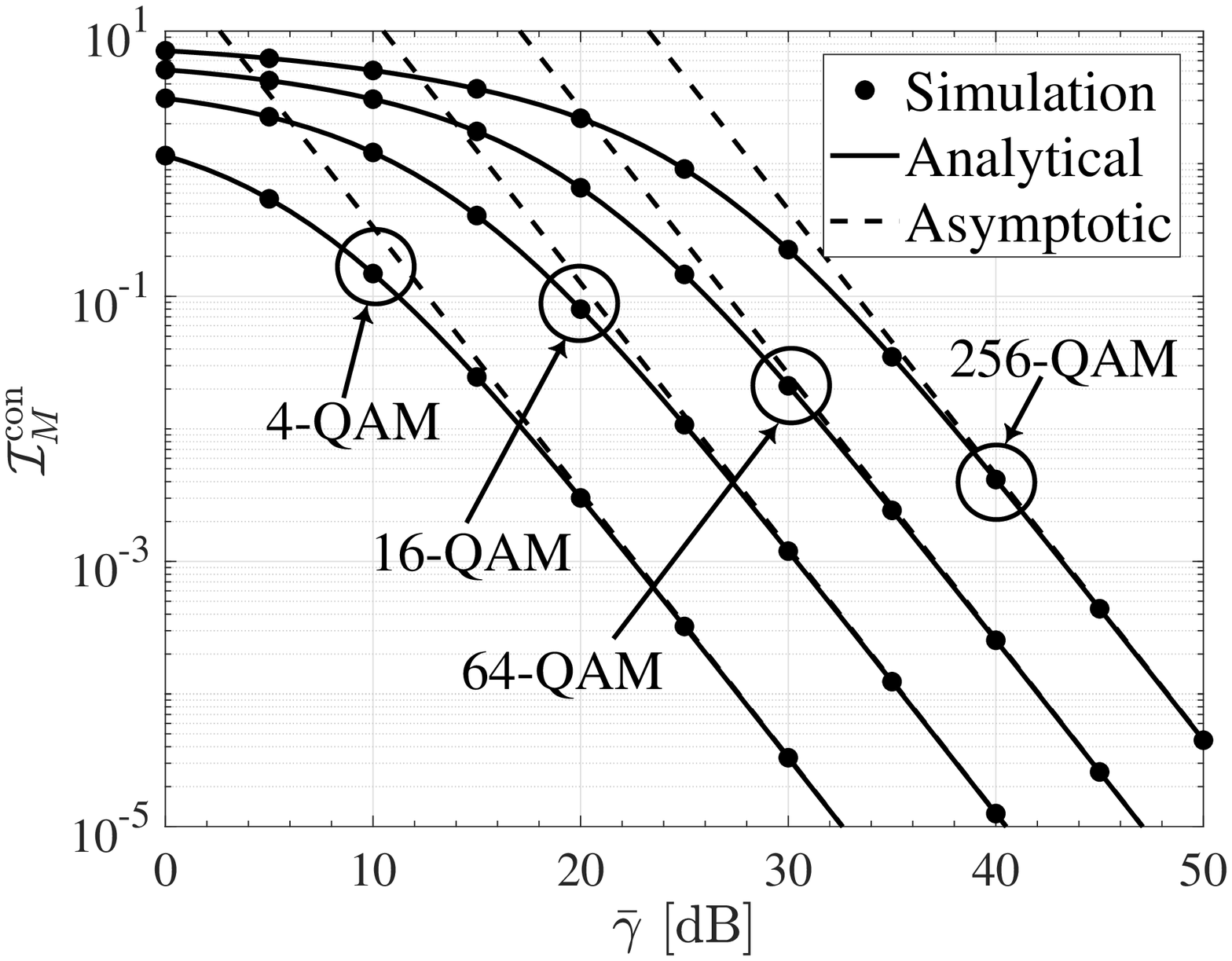}
	   \label{fig1b}	
    }\\\vspace{0pt}
    \subfigure[$\kappa$-$\mu$, $\kappa=1$, $\mu=2$.]
    {
        \includegraphics[height=0.185\textwidth]{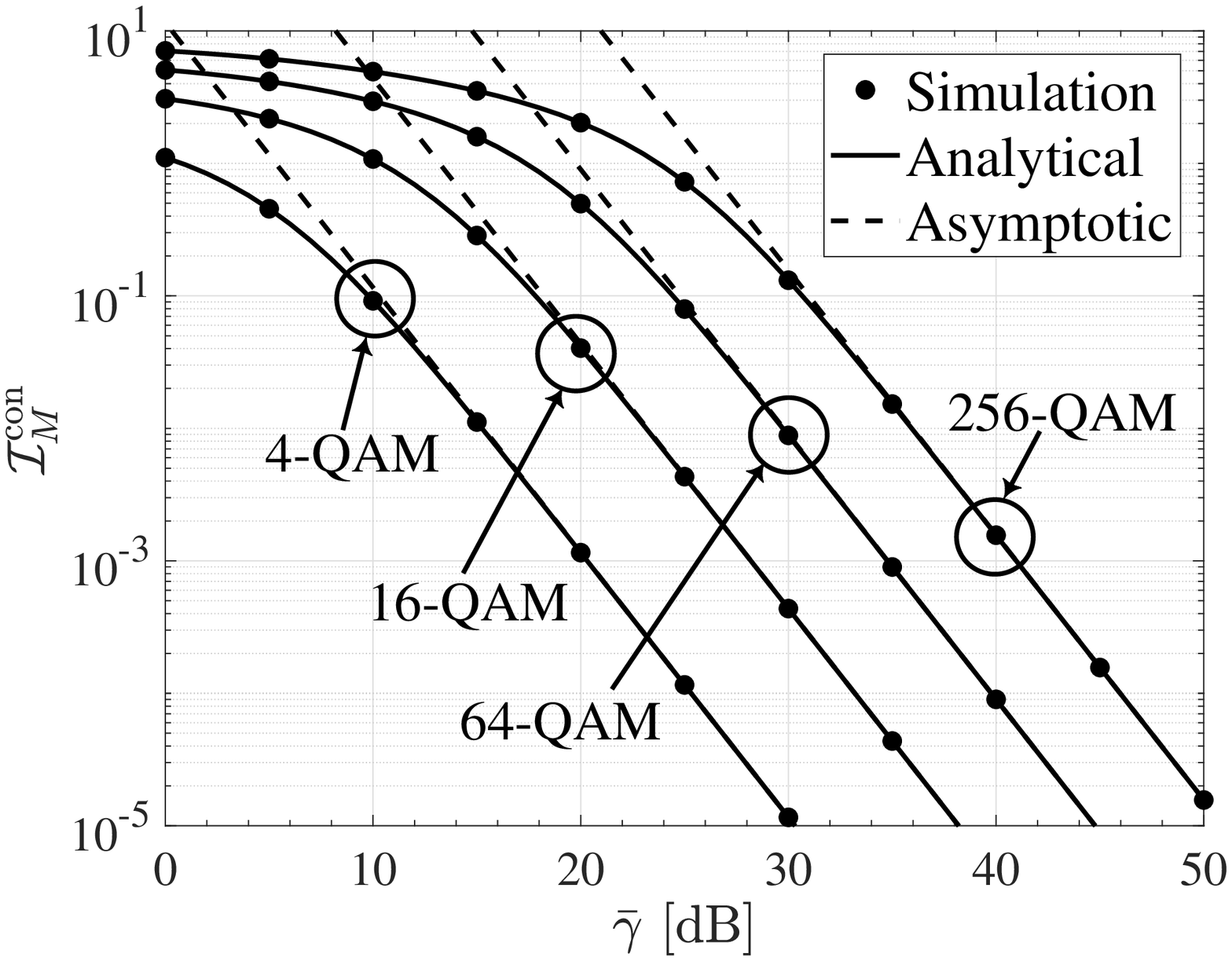}
	   \label{fig1c}	
    }
    \hspace{-23pt}
   \subfigure[${\mathcal K}_G$, $k=4$, $m=2$, $L=30$.]
    {
        \includegraphics[height=0.185\textwidth]{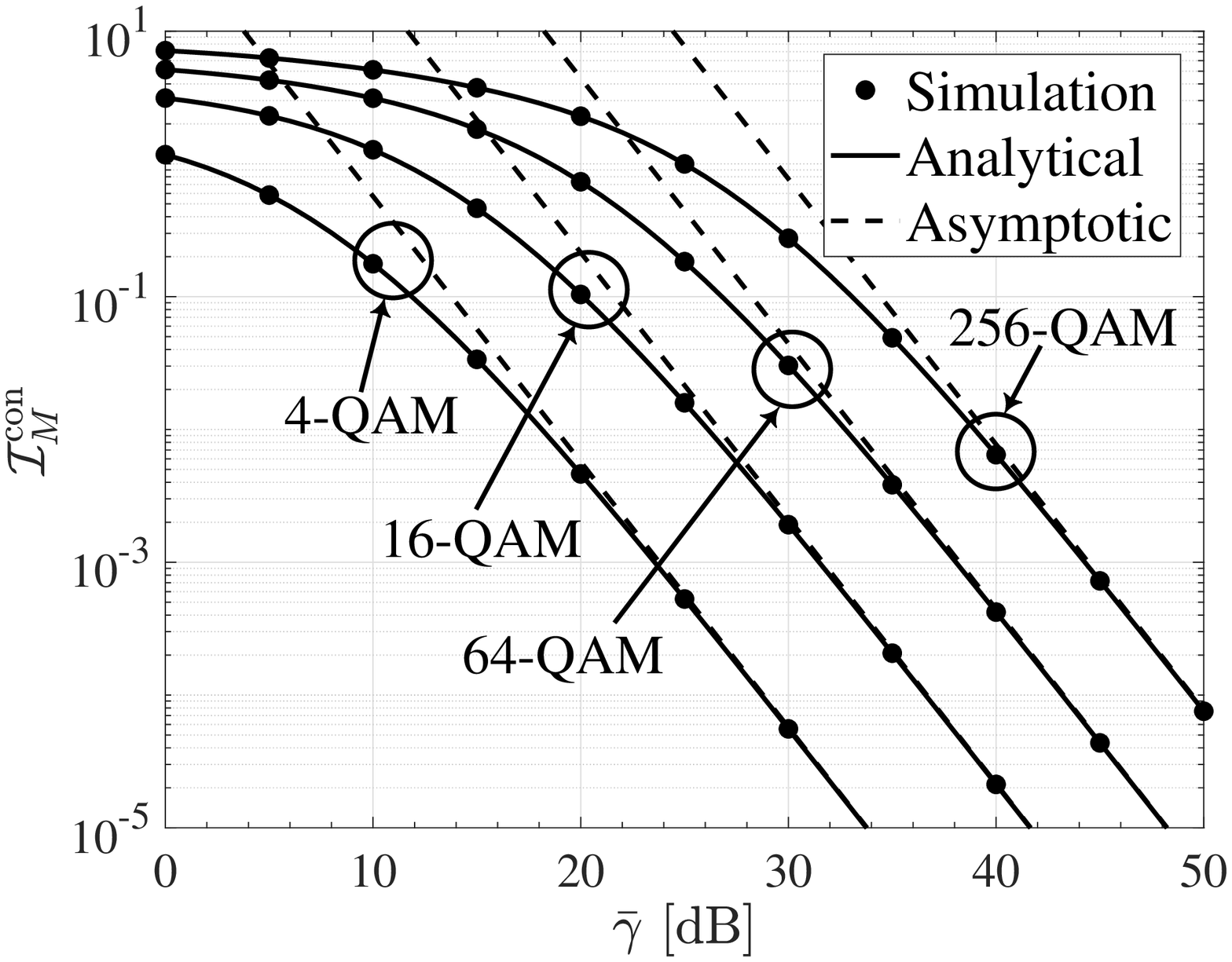}
	   \label{fig1d}	
    }\\
\caption{Asymptotic AMI of $M$-QAM versus $\bar\gamma$ over different fading channels.}
    \label{figure1}
    \vspace{-15pt}
\end{figure}

\begin{figure}[!t]
    \centering
    \subfigbottomskip=0pt
	\subfigcapskip=-5pt
\setlength{\abovecaptionskip}{0pt}
    \subfigure[Nakagami-$m$ channels.]
    {
        \includegraphics[height=0.18\textwidth]{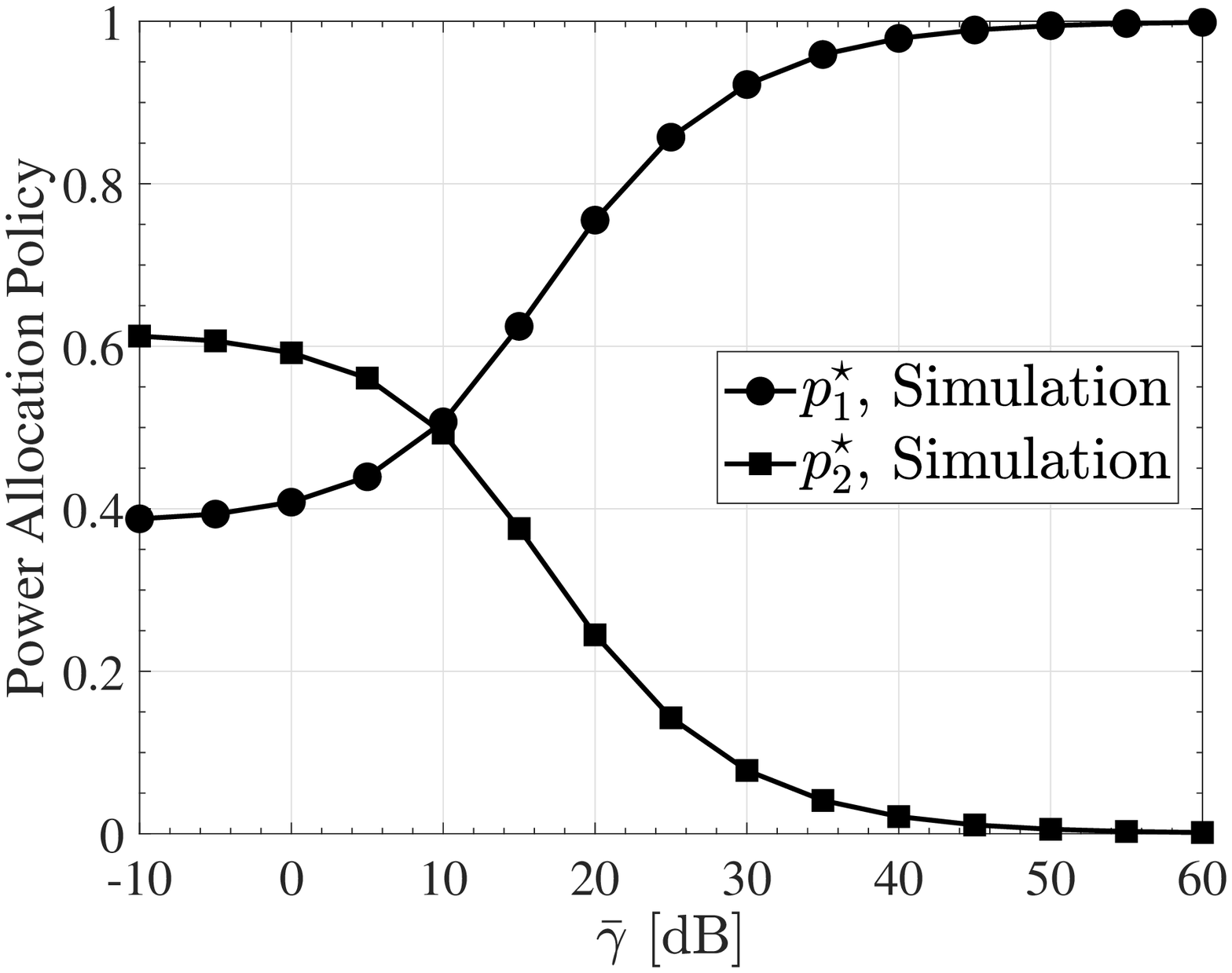}
	   \label{fig2a}	
    }
    \hspace{-12pt}
   \subfigure[$\kappa$-$\mu$ channels.]
    {
        \includegraphics[height=0.18\textwidth]{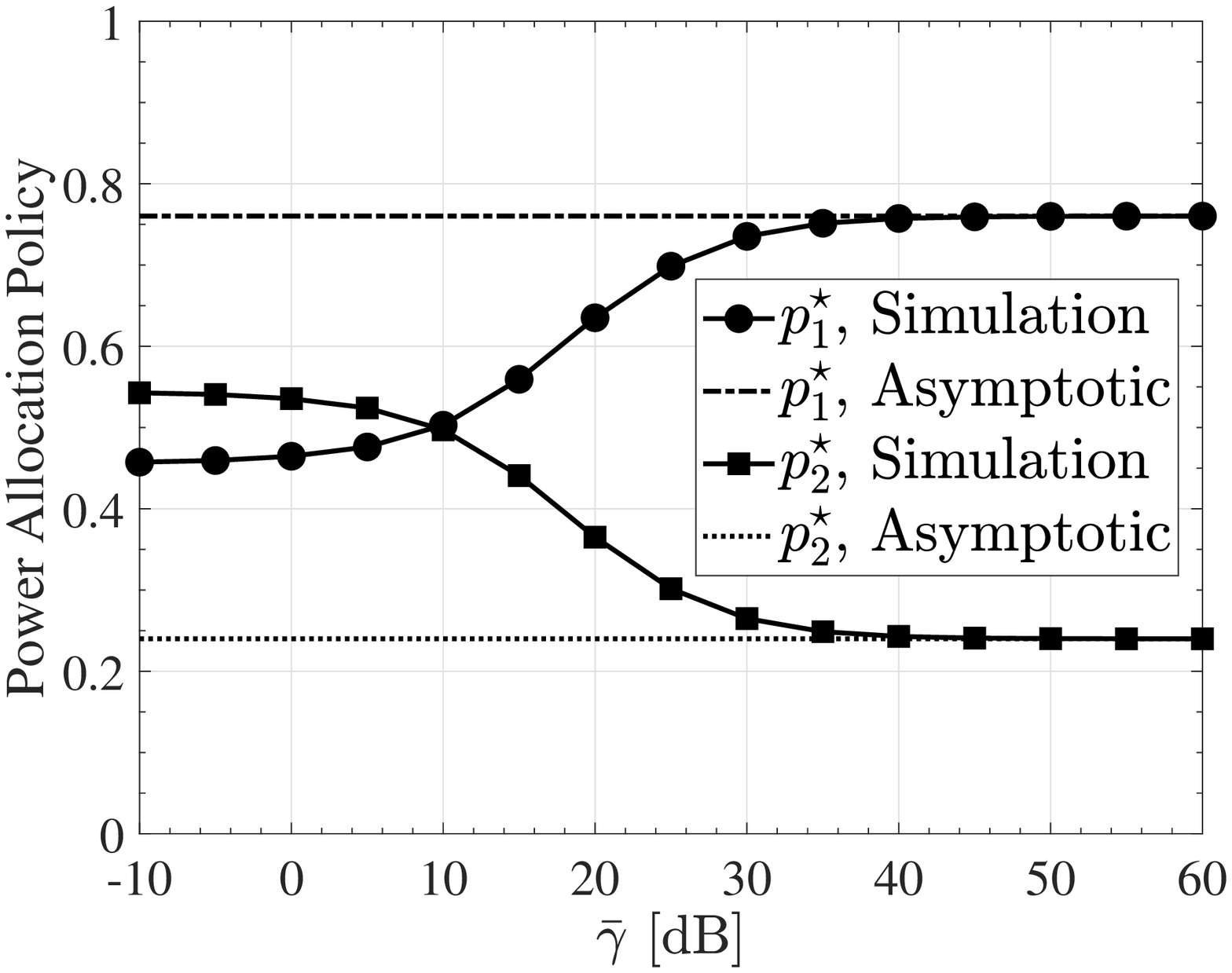}
	   \label{fig2b}	
    }
\caption{Optimal power allocation policy in a bank of two parallel fading channels having 4-QAM inputs: {\figurename} {\ref{fig2a}}, Nakagami-$m$ fading ($m_1=1$ and $m_2=4$); {\figurename} {\ref{fig2b}}, $\kappa$-$\mu$ fading ($\kappa_1=2$, $\mu_1=1$, $\kappa_2=5$, and $\mu_2=1$).}
    \label{figure2}
    \vspace{-15pt}
\end{figure}

Numerical simulations are performed to verify the rightness of the theoretical analyses. {\figurename} {\ref{figure1}} plots ${\mathcal I}_{M}^{\text {con}}=\log_2{M}-{\bar{{\mathcal I}}}_{M}$ versus $\bar\gamma$ over different mixture gamma distributed channels to validate the correctness of \eqref{Asym_AMI_Case5}. The analytical results are obtained with the aid of \eqref{AMI_Appr}, where $N$ is set as 30, whereas the asymptotic results are calculated by $\log_2{M}-{\bar{{\mathcal I}}}_{M}^{\infty}=\frac{G_{a,M}}{\bar\gamma^{G_{d,M}}}$. Note that $I_M^{\mathcal X}\left(\cdot\right)$ in \eqref{AMI_Appr} and $G_{a,M}$ can be calculated by the method of numerical integration. As shown, in the high SNR regime, the derived asymptotic results track the numerical results accurately. Then, we use {\figurename} {\ref{figure2}} to illustrate the optimal power allocation policy in a bank of two parallel independent MG distributed fading channels, where the simulated $p_i^{\star}$ is obtained by the method introduced in \cite{Lozano2008}. The scenario of Nakagami-$m$ fading is plotted in {\figurename} {\ref{fig2a}}, where the fading parameters for sub-channels 1 and 2 are $m_1=1$ and $m_2=4$, respectively. As previously mentioned, the diversity order of Nakagami-$m$ channel is $m$. Due to this, sub-channel 1 yields a higher diversity order than sub-channel 2, and thus we have $p_{1}^{\star}\rightarrow1$ and $p_2^{\star}\rightarrow0$ as $\bar\gamma\rightarrow\infty$, which is confirmed by {\figurename} {\ref{fig2a}}. {\figurename} {\ref{fig2b}} illustrates the case of $\kappa$-$\mu$ fading, where the fading parameters for sub-channels 1 and 2 are ($\kappa_1=2$, $\mu_1=1$) and ($\kappa_2=5$, $\mu_2=1$), respectively. Note that $\mu_1=\mu_2$, indicating these two sub-channels yield the same diversity order. In this case, the value of $p_i^{\star}$ for $\bar\gamma\rightarrow\infty$ can be calculated by \eqref{Power_Allocate_Prop2}. As can be seen from {\figurename} {\ref{fig2b}}, in the high SNR regime, $p_i^{\star}$ converges to the value calculated by \eqref{Power_Allocate_Prop2}, which thus validates the rationality of our analyses.

\section{Conclusion}
\label{section4}
A unified framework was established to analyze the asymptotic AMI over MG distributed fading channels with finite inputs. The analytical results were further used to discuss the asymptotic optimal power allocation policy of parallel fading channels, suggesting more power should be allocated to the sub-channel with a lower coding gain and/or diversity order.

\begin{appendices}
\section{Proof of Theorem \ref{theorem2}}
\label{Append0}
\begin{IEEEproof}
By \cite{Guo2005}, $\lim_{t\rightarrow0^{+}}{\text{mmse}}_{M}^{\mathcal X}\left(t\right)=1$. Moreover, by Lemma \ref{theorem1}, $\lim_{t\rightarrow\infty}{\text{mmse}}_{M}^{\mathcal X}\left(t\right)=o\left({\rm e}^{-\frac{t}{8}d_{{\mathcal X},{\min}}^2}\right)$, where $o\left(\cdot\right)$ denotes the higher order term. In summary, ${\text{mmse}}_{M}^{\mathcal X}\left(t\right)$ is ${\mathcal O}\left(t^0\right)$ as $t\rightarrow0^{+}$ and ${\mathcal O}\left(t^{-\infty}\right)$ as $t\rightarrow\infty$, which together with the fact of $\beta_l>0$ \cite{Tellambura2011} and Lemma \ref{theorem0}, yields $\hat{\mathcal M}\left({\beta_l}\right)<\infty$ and $\hat{\mathcal M}\left({\beta_l}+1\right)<\infty$. Additionally, based on \cite{Tellambura2011}, we find $\beta_{l+1}-\beta_{l}=0$ or $\beta_{l+1}-\beta_l\geq 1$ holds, and thus \eqref{Asym_AMI_Case5} can be simplified to ${\bar{{\mathcal I}}}_{M}^{\infty}=\log_2{M}-G_{a,M}\bar\gamma^{-\beta_1}+
{\mathcal O}\left({\bar\gamma^{-\beta_1-1}}\right)$. This completes the proof of Theorem \ref{theorem2}.
\end{IEEEproof}
\section{Proof of Theorem \ref{lemma1}}
\label{Append1}
\begin{IEEEproof}
Leveraging the Karush–Kuhn–Tucker (KKT) optimality conditions of $P_0$, we obtain
{\setlength\abovedisplayskip{1pt}
\setlength\belowdisplayskip{1pt}
\begin{align}\label{Allocate_Cond}
\left.-\frac{\partial{\mathcal{S}}_M^{\mathcal X}\left(\textbf p\right)}{\partial p_k}\right|_{p_k=p_k^{\star}}\left\{
\begin{array}{rcl}
\leq \nu       &      & {p_k^{\star}=0}\\
=\nu     &      & {p_k^{\star}>0}
\end{array}, \right.
\end{align}
}where $\nu$ is such that $\sum_{k=1}^{K}p_k^{\star}=1$. It can be derived from $\frac{{\rm d}I_{M}^{\mathcal X}\left(\gamma\right)}{{\rm d}\gamma}={\textrm{mmse}}_{M}^{\mathcal X}\left(\gamma\right)$ \cite{Guo2005} that
\begin{align}
-\frac{\partial{\mathcal{S}}_M^{\mathcal X}\left(\textbf p\right)}{\partial p_k}=&{\mathbb E}\left\{\frac{{\rm d}{I}_M^{\mathcal X}\left(a_kp_k\bar\gamma \right)}{{\rm d}p_k}\right\}=\bar\gamma{\mathbb E}\left\{{\textrm{mmse}}_M^{\mathcal X}\left(a_kp_k\bar\gamma \right)a_k\right\}\nonumber\\
=&\bar\gamma\int_{0}^{\infty}xf_k\left(x\right){\textrm{mmse}}_{M}^{\mathcal X}\left(\bar\gamma p_k x\right){\rm d}x,\label{Dev_3}
\end{align}
which, together with the fact of ${\textrm{mmse}}_M^{\mathcal X}\left(0 \right)=1$ \cite{Guo2005}, yields $\left.-\frac{\partial{\mathcal{S}}_M^{\mathcal X}\left(\textbf p\right)}{\partial p_k}\right|_{p_k=0}=\bar\gamma\int_{0}^{\infty}xf_k\left(x\right){\rm d}x=\bar\gamma{\mathbb E}\left\{a_k\right\}=\bar\gamma$ and $\bar\gamma{\mathbb E}\left\{{\textrm{mmse}}_M^{\mathcal X}\left(a_kp_k^{\star}\bar\gamma \right)a_k\right\}=\nu$ ($p_k^{\star}>0$). Furthermore, it follows, from \eqref{AMI_Def}, that
\begin{align}\label{AMI_Asym_Final3}
\frac{{\rm d}{{\bar{\mathcal I}}_M}}{{\rm d}\bar\gamma}=\int_{0}^{\infty}af\left(a\right){\textrm{mmse}}_{M}^{\mathcal X}\left(\bar\gamma a\right){\rm d}a.
\end{align}
As a result of \eqref{AMI_Asym_Final2_Appl} and \eqref{AMI_Asym_Final3}, for $p_k^{\star}>0$ and $\bar\gamma\rightarrow\infty$, we have
\begin{align}
\nu{\bar\gamma^{D_{M,k}}}=&\bar\gamma^{D_{M,k}}\bar\gamma{\mathbb E}\left\{{\textrm{mmse}}_M^{\mathcal X}\left(a_kp_k^{\star}\bar\gamma \right)a_k\right\}\\
=&{A_{M,k}D_{M,k}}{{p_k^{\star}}^{-D_{M,k}-1}}+{\mathcal O}\left({\bar\gamma^{-1}}\right)\label{Asym_Power_Result1}
\end{align}
and $\lim_{\bar\gamma\rightarrow\infty}\nu=\lim_{\bar\gamma\rightarrow\infty}\frac{\bar\gamma A_{M,k}D_{M,k}}{\left(p_k^{\star}\bar\gamma\right)^{D_{M,k}+1}}=0$, which suggests that $\left.-\frac{\partial{\mathcal{S}}_M^{\mathcal X}\left(\textbf p\right)}{\partial p_k}\right|_{p_k=0}=\bar\gamma>\nu$ holds for $\bar\gamma\rightarrow\infty$. Taken this and \eqref{Allocate_Cond} together, it can be concluded that the limiting optimal power allocation is strictly positive, namely $\lim_{\bar\gamma\rightarrow\infty}p_k^{\star}>0$, $\forall k\in\mathcal K$. Besides, by \eqref{Asym_Power_Result1}, we find that for any $k\in\mathcal K$ that satisfies $D_{M,k}=D_{\min}$, it has
\begin{align}\label{Power_Allocate_Prop1}
\nu{\bar\gamma^{D_{M,k}}}=\nu{\bar\gamma^{D_{\min}}}={\mathcal O}\left(1\right),\quad\bar\gamma\rightarrow\infty,
\end{align}
which leads to
\begin{align}
&p_k^{\star}={\left(\frac{A_{M,k}D_{\min}}{\nu{\bar\gamma^{D_{\min}}}+{\mathcal O}\left({\bar\gamma^{-1}}\right)}\right)}^{\frac{1}{D_{\min}+1}}\nonumber\\
&={\left(\frac{A_{M,k}D_{\min}}{\nu{\bar\gamma^{D_{\min}}}}\right)}^{\frac{1}{D_{\min}+1}}
{\left(\frac{1}{1+{\mathcal O}\left({\bar\gamma^{-1}}\right)\nu^{-1}{\bar\gamma^{-D_{\min}}}}\right)}^{\frac{1}{D_{\min}+1}}\nonumber\\
&={\left(\frac{A_{M,k}D_{\min}}{\nu{\bar\gamma^{D_{\min}}}}\right)}^{\frac{1}{D_{\min}+1}}+{\mathcal O}\left({\bar\gamma^{-1}}\right),~\bar\gamma\rightarrow\infty.\label{Asym_Power_Der2}
\end{align}
Here, \eqref{Asym_Power_Der2} is due to the fact of $\lim_{x\rightarrow0}\left(\frac{1}{1+x}\right)^{m}=1+\mathcal{O}\left(x\right)$, $m>0$. As for $k\in\mathcal K$ that satisfies $D_{M,k}>D_{\min}$, we have
\begin{align}
p_k^{\star}=&{\left(\frac{A_{M,k}D_{M,k}}{\nu{\bar\gamma^{D_{\min}}}\bar\gamma^{D_{M,k}-D_{\min}}+{\mathcal O}\left({\bar\gamma^{-1}}\right)}\right)}^{\frac{1}{D_{M,k}+1}}\nonumber\\
=&{\left(\frac{A_{M,k}D_{M,k}}{\nu{\bar\gamma^{D_{\min}}}}\right)}^{\frac{1}{D_{M,k}+1}}{\bar\gamma^{\frac{D_{\min}-D_{M,k}}{D_{M,k}+1}}}\nonumber\\
&+{\mathcal O}\left({\bar\gamma^{\left(D_{\min}-D_{M,k}\right){\frac{D_{M,k}+2}{D_{M,k}+1}}-1}}\right),~\bar\gamma\rightarrow\infty,\label{Asym_Power_Der3}
\end{align}
where the second equality is also based on the fact of $\lim_{x\rightarrow0}\left(\frac{1}{1+x}\right)^{m}=1+\mathcal{O}\left(x\right)$, $m>0$. Taken together, in the high SNR regime, the optimal power allocation policy obeys \eqref{Asym_Power_Der2} and \eqref{Asym_Power_Der3}, where $\nu$ is such that $\sum_{k=1}^{K}p_k^{\star}=1$. This completes the proof of Theorem \ref{lemma1}.
\end{IEEEproof}
\end{appendices}

\end{document}